\documentclass{PoS}

\usepackage[british]{babel}
\usepackage[T1]{fontenc}				
\usepackage[utf8]{inputenc}				

\usepackage{graphicx}

\usepackage{mathrsfs}
\usepackage{amssymb}
\usepackage{amsmath}
\usepackage{mathtools}

\usepackage{cite}
\usepackage{mciteplus}

\def\figscale{0.36}
\def\figspace{0mm}

\newcommand{\xt}{{\mathbf{x}}}

\newcommand{\ktt}{k_\perp} 

\newcommand{\ud}{\mathrm{d}}

\newcommand{\tr}{\, \mathrm{Tr} \, }

\newcommand{\nc}{{N_\mathrm{c}}}

\newcommand{\cf}{C_\mathrm{F}}

\newcommand{\nr}[1]{(\ref{#1})}

\newcommand{\as}{\alpha_{\mathrm{s}}}

\newcommand{\xbj}{{x_{Bj}}}

\newcommand{\kcal}{\mathcal{K}}

\title{
	Factorization of the soft gluon divergence from the dipole picture deep inelastic scattering cross sections at next-to-leading order
}

\ShortTitle{Dipole Picture NLO DIS}

\author{Bertrand Ducloué\\
	Institut de Physique Théorique, Université Paris-Saclay, CEA, CNRS, F-91191 Gif-sur-Yvette, France\\
	E-mail: \email{bertrand.ducloue@ipht.fr}
}

\author{\speaker{Henri Hänninen}\\
	Department of Physics, P.O. Box 35, 40014 University of Jyv\"askyl\"a, Finland\\
	E-mail: \email{henri.j.hanninen@student.jyu.fi}
}
   
\author{Tuomas Lappi\\
	Department of Physics, P.O. Box 35, 40014 University of Jyv\"askyl\"a, Finland\\
	Helsinki Institute of Physics, P.O. Box 64, 00014 University of Helsinki, Finland\\
	E-mail: \email{tuomas.v.v.lappi@jyu.fi}
}

\author{Yan Zhu\\
	Physik-Department, Technische Universität München, D-85748 Garching, Germany\\
	E-mail: \email{yanzhu.zhu@tum.de}
}

\abstract{

We use a factorization scheme analogous to one proposed for single inclusive forward hadron production to factorize the soft gluon divergence present in the deep inelastic scattering cross sections in the dipole picture at next-to-leading order (NLO). We show numerically that in this carefully constructed scheme it is possible to obtain meaningful results for the DIS cross sections at NLO, and so we are able to quantitatively study the recently derived NLO corrections to the DIS cross sections. We find that the NLO corrections can be significant and sensitive to the details of the factorization scheme used for the resummation of the large logarithms into the BK evolution equation. In the case of an approximative factorization scheme we observe a problematic behavior of the DIS cross sections similar to what has been seen with analogously factorized single inclusive cross sections.
}

\FullConference{XXVI International Workshop on Deep-Inelastic Scattering and Related Subjects (DIS2018)\\
		16-20 April 2018\\
		Kobe, Japan}

\begin{document}

\section{Introduction}
Deep inelastic scattering (DIS) provides a clean process to study the partonic structure of hadrons. At small Bjorken-$x$ it is convenient to look at the process in the dipole picture where the scattering factorizes into two parts: first the virtual photon fluctuates into a quark-antiquark pair in a QED process, and subsequently the quark dipole scatters off the target in a QCD process. Already at leading order the dipole picture has led to satisfactory fits to HERA DIS data, using the running coupling BK equation \cite{Balitsky:1995ub,Kovchegov:1999yj}, see e.g. Refs. \cite{Albacete:2010sy,Lappi:2013zma}. Recent progress on both NLO BK \cite{Balitsky:2008zza,Beuf:2014uia,Iancu:2015vea,Iancu:2015joa,Lappi:2016fmu} and NLO DIS impact factors \cite{Beuf:2016wdz,Beuf:2017bpd,Hanninen:2017ddy} have made full NLO cross section computations possible in the dipole picture.

In this work we construct a subtraction scheme for the resummation of the large logarithms of energy present in the NLO DIS impact factors according to the principle presented in Ref. \cite{Iancu:2016vyg}, which was shown to be effective in the case of single inclusive particle production in Ref. \cite{Ducloue:2017mpb}. To demonstrate the effectiveness of the subtraction procedure we 
computed in Ref. \cite{Ducloue:2017ftk} the DIS structure functions at NLO accuracy.
The numerical results allow us to evaluate the importance of the NLO contributions and to estimate the stability of the perturbative expansion for this quantity.

\section{Next-to-leading order cross sections and soft gluon divergence}

The total photon-proton cross sections at leading order in the dipole picture for a transversely (\textit{T}) and longitudinally (\textit{L}) polarized photon read
\begin{align}\label{eq:lo}
\sigma_{L,T}^{\text{LO}}(\xbj,Q^2) & = 4 \nc \alpha_{em} \sum_f e_f^2 \int_0^1 \ud z_1 
\int_{\xt_0, \xt_1} \kcal_{L,T}^{\text{LO}}(z_1,\xt_0,\xt_1,\xbj),
\end{align}
with the notation $\int_{\xt_0} =\int\frac{\ud^2 \xt_0}{2\pi}$. The integrands above are products of the light cone wavefunctions for the $\gamma^{*} \rightarrow q \bar{q}$ fluctuation and the $q \bar{q}$ dipole--color field target scattering amplitudes:
\begin{align}
\kcal_{L}^{\text{LO}}(z_1,\xt_0,\xt_1,X) &= 4 Q^2 z_1^2 (1-z_1)^2 
K_0^2(Q X_2) \left(1-S_{01}(X)\right), \\
\kcal_{T}^{\text{LO}}(z_1,\xt_0,\xt_1,X) &= Q^2 z_1 (1-z_1) \left(z_1^2+(1-z_1)^2\right)
K_1^2(Q X_2) \left(1-S_{01}(X)\right),
\end{align}
where $X_2^2 = z_1 (1-z_1) \xt_{01}^2$.
The scattering matrix $S_{01}$ for the dipole -- color field scattering is given by the two-point correlation function of Wilson lines:
\begin{align}
S_{01}(X) \equiv S(\xt_{01}=\xt_0-\xt_1,X) 
=\left< \frac{1}{\nc}\tr U(\xt_0)U^\dag(\xt_1) \right>_X,
\end{align}
where $X$ is the momentum fraction at which the two-point correlation function is evaluated. The momentum fraction is related to the BK equation evolution variable via $y = \ln 1/X$.

At next-to-leading order the virtual photon Fock state contains contributions from a gluon loop to the quark-antiquark dipole and a new parton state $q \bar{q} g$ with a quark-antiquark-gluon tripole, where in the former case the $q \bar{q}$ dipole and in the latter the $q \bar{q} g$ tripole scatters off the target. These NLO corrections have been calculated in mixed space by G. Beuf~\cite{Beuf:2016wdz,Beuf:2017bpd} using conventional dimensional regularization,
and verified in the four-dimensional helicity scheme \cite{Hanninen:2017ddy}. The total $\gamma^* p$ scattering cross section at next-to-leading order can be written in the following "unsubtracted" form in accordance with the general idea presented in Ref. \cite{Iancu:2016vyg}:
\begin{equation}
\label{eq:NLO_bare}
\sigma_{L,T}^{\text{NLO}}
=\sigma_{L,T}^{\text{IC}}
+\sigma_{L,T}^{qg}
+\sigma_{L,T}^{\text{dip}} \, .
\end{equation}
Here the first term is the lowest order contribution with an unevolved target
and the last two terms contain all the corrections proportional to $\alpha_s$. The quark-gluon $\sigma_{L,T}^{qg}$ and dipole $\sigma_{L,T}^{\text{dip}}$ contributions can be thought to emerge from the computation of the NLO real emission and gluon loop diagrams, respectively. However, the diagram computation results are separately UV divergent and so must be combined in a way that shows the cancellation of the divergences, which has been done for Eq. \eqref{eq:NLO_bare}. This was done in Ref. \cite{Beuf:2017bpd} by introducing suitable subtraction terms. The choice of the subtraction terms is not unique and the UV subtraction was done in an alternative way in Ref. \cite{Hanninen:2017ddy} yielding equivalent final results. We have written the UV finite results from Ref. \cite{Beuf:2017bpd} in the following form:
\begin{align} \label{eq:NLO_qg_bare}
&\sigma_{L,T}^{qg}
= 8 \nc \alpha_{em} \frac{\alpha_s \cf}{\pi} \sum_f e_f^2 \int_0^1 \ud z_1 \int^{1-z_1} \frac{\ud z_2}{z_2} 
\int_{\xt_0, \xt_1, \xt_2} \!\! \kcal_{L,T}^{\text{NLO}}\left(z_1,z_2,\xt_0,\xt_1,\xt_2,X(z_2)\right) , 
\\
&\sigma_{L,T}^{\text{dip}}
= 4 \nc \alpha_{em} \frac{\alpha_s \cf}{\pi} \sum_f e_f^2 \int_0^1 \ud z_1 
\int_{\xt_0, \xt_1} \!\! \kcal_{L,T}^{\text{LO}}(z_1,\xt_0,\xt_1,X^{\text{dip}}) \! \left[\frac{1}{2}\ln^2\!\left(\!\frac{z_1}{1\!-\!z_1}\!\right)\!-\!\frac{\pi^2}{6}\!+\!\frac{5}{2}\right],
\label{eq:NLO_dip}
\end{align}
where the expressions for the NLO integrands $\kcal_{L,T}^{\text{NLO}}$ can be found in Ref. \cite{Ducloue:2017ftk}. 
In the above $\xt_0, \xt_1, \xt_2$ are the transverse positions of the quark, antiquark and gluon, $z_i$ their longitudinal momentum fractions, and the $q \bar{q} g$ state Wilson line scattering operator is
\begin{align}
S_{012}(X) & =\frac{\nc}{2\cf}\left(S_{02}(X)S_{21}(X)-\frac{1}{\nc^2}S_{01}(X)\right).
\end{align}
Note the incompletely defined logarithmic integral over $z_2$ in Eq. \eqref{eq:NLO_qg_bare}. Since the integrands $\kcal_{L,T}^{\text{NLO}}\left(z_1,z_2,\xt_0,\xt_1,\xt_2,X\right)$ tend to non-zero values as $z_2 \to 0$ at fixed $X$, the $qg$ contributions are logarithmically divergent. This large logarithm needs to be subtracted and resummed into the target BK evolution. On the other hand the "dipole"-contribution \eqref{eq:NLO_dip} does not contain such a large logarithm and so has been integrated over the internal gluon loop momentum fraction $z_2$.

We will briefly review the subtraction procedure constructed in Ref. \cite{Ducloue:2017ftk}. First the lowest order term in Eq. \eqref{eq:NLO_bare} is identified as the leading order cross section \eqref{eq:lo} without leading log resummation, i.e. with the dipole scattering amplitude evaluated at the initial condition $X=x_0$.
In order to achieve a stable perturbative expansion at NLO the dipole amplitudes in the $qg$-contribution $\sigma_{L,T}^{qg}$ must be evaluated at a rapidity that depends on the fractional momentum $z_2$ of the emitted gluon\cite{Iancu:2016vyg,Beuf:2014uia}.
We argue from kinematics that since at small $z_2$ the target momentum fraction behaves as $X(z_2) \approx \ktt^2/(z_2 W^2)$, where $\ktt$ is the gluon transverse momentum, that it might be possible to approximate $X(z_2) \approx \xbj/z_2$ for DIS, with similar argumentation as for the single inclusive particle production in Ref. \cite{Iancu:2016vyg,Ducloue:2017mpb}. At small target momentum fraction $X < x_0$ this yields a lower limit $z_2 > \xbj / x_0$.

Now we can complete the "unsubtracted" form of the NLO cross sections \eqref{eq:NLO_bare} by setting $X(z_2) \equiv \xbj/z_2$ and the limit $z_2 > \xbj / x_0$.
It might be preferable to write the NLO cross section as the full leading order cross section \eqref{eq:lo} and some $\alpha_s$ corrections. To this end, we note that by taking the $z_2 \to 0$ limit of $\kcal_{L,T}^{\text{NLO}}$ (in the explicit $z_2$ dependence, not in the implicit through $X(z_2)$) one gets an integral version of the BK evolution equation. Using this fact we write the "subtracted" form of the NLO cross section:
\begin{equation}
\label{eq:NLO_sub}
\sigma_{L,T}^{\text{NLO}}
=\sigma_{L,T}^{\text{LO}}
+\sigma_{L,T}^{qg, \text{sub.}}
+\sigma_{L,T}^{\text{dip}} \, ,
\end{equation}
where $\sigma_{L,T}^{\text{LO}}$ is the leading order expression \nr{eq:lo} with evolved target and
\begin{align}
\label{eq:NLO_qg_sub}
\sigma_{L,T}^{qg, \text{sub.}}
= 
& 8 \nc \alpha_{em} \frac{\alpha_s \cf}{\pi} \sum_f e_f^2  \int_0^1 \ud z_1 \int_{\xbj/x_0}^{1} \frac{\ud z_2}{z_2} 
\int_{\xt_0, \xt_1, \xt_2} \! 
\bigg[ \theta(1\!-\!z_1\!-\!z_2)
\nonumber \\ & \times
\kcal_{L,T}^{\text{NLO}}\left(z_1,z_2,\xt_0,\xt_1,\xt_2,X(z_2)\right) 
- \kcal_{L,T}^{\text{NLO}}\left(z_1,0,\xt_0,\xt_1,\xt_2,X(z_2)\right)\bigg] .
\end{align}

It might be tempting to make simplifying approximations to the subtracted scheme by neglecting the lower limit in $z_2$ as small and the $z_2$ dependence in the target momentum fraction $X(z_2)$. With these adjustments one gets the "$\xbj$-subtracted" scheme:
\begin{equation}
\label{eq:NLO_xbjsub}
\sigma_{L,T}^{{\text{NLO}},\xbj-\text{sub.}}
=\sigma_{L,T}^{\text{LO}}
+\sigma_{L,T}^{qg, \text{sub.*}}
+\sigma_{L,T}^{\text{dip}} \, ,
\end{equation}
where in $\sigma_{L,T}^{qg, \text{sub.*}}$ we take $X(z_2)=\xbj$ and $\xbj/x_0 \to 0$. This $\xbj$-subtracted scheme is analogous to the "CXY" scheme used in single inclusive particle production in \cite{Ducloue:2017mpb} and while this approximative scheme is formally equivalent at this order of perturbation theory, the "CXY" scheme has been shown to lead to problematic results at high momentum scales.

\section{Numerical results}

To demonstrate the behavior of the unsubtracted and $\xbj$-subtracted schemes we present some of our results from Ref. \cite{Ducloue:2017ftk}. We neglect impact parameter effects 
so the presented plots are of $F_{L,T}/(\sigma_0/2)$ where the structure functions are
$
F_{L,T}(\xbj,Q^2)=\frac{Q^2}{4\pi^2\alpha_{\text{em}}}\sigma_{L,T}(\xbj,Q^2).
$
For the target evolution the LO BK equation with an MV initial condition~\cite{McLerran:1993ni} was used. To study the importance  of running coupling effects, a fixed coupling $\alpha_{\mathrm{s}} = 0.2$ was compared to parent dipole running coupling $\alpha_{\mathrm{s}} = \alpha_{\mathrm{s}}(\xt_{01}^2) = \alpha_{\mathrm{s}} (Q^2 = 4C^2/\xt_{01}^2)$.	


\begin{figure*}[tbp]
	\centering
	\includegraphics[scale=\figscale]{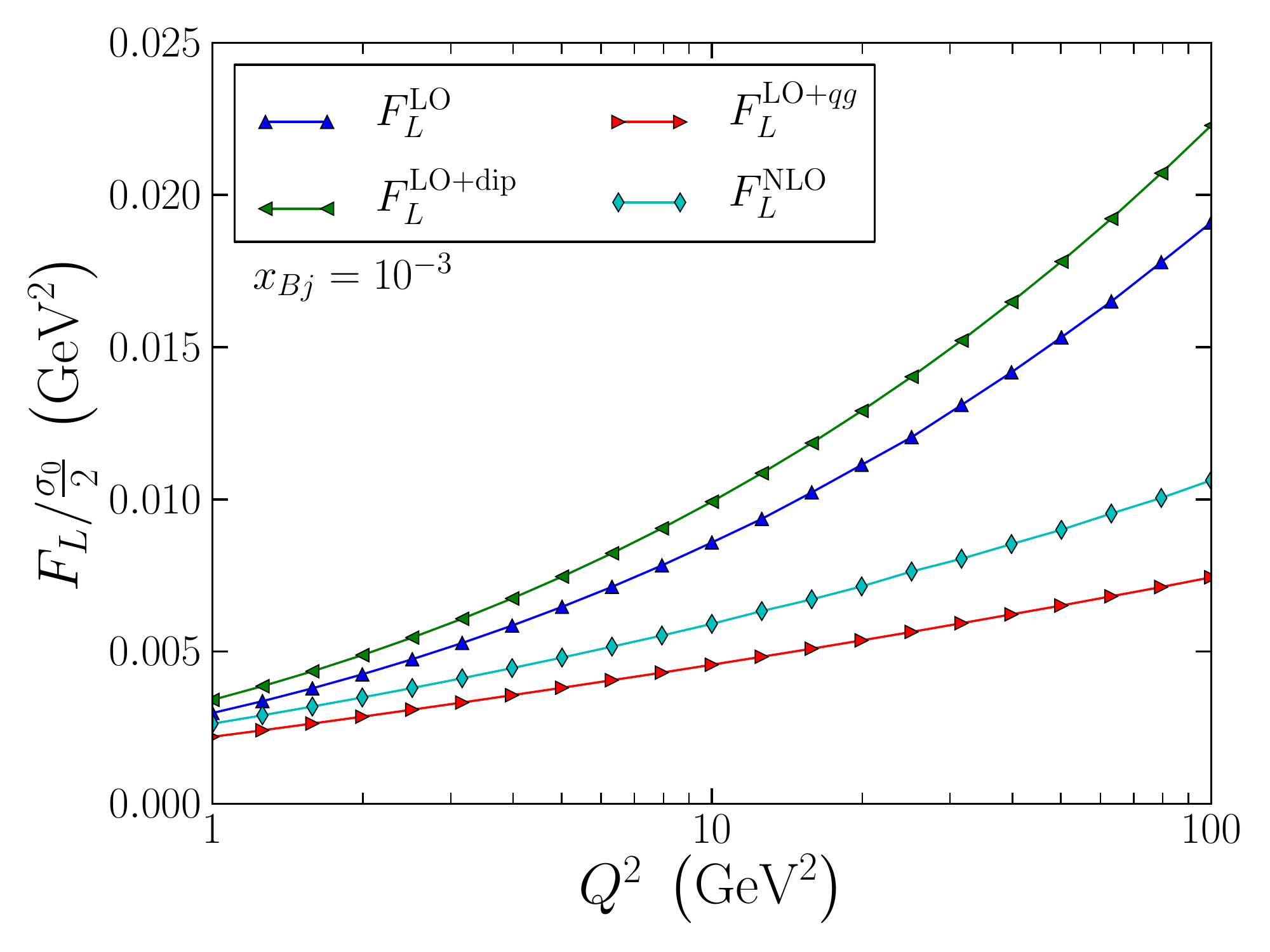}
	\hspace{\figspace}
	\includegraphics[scale=\figscale]{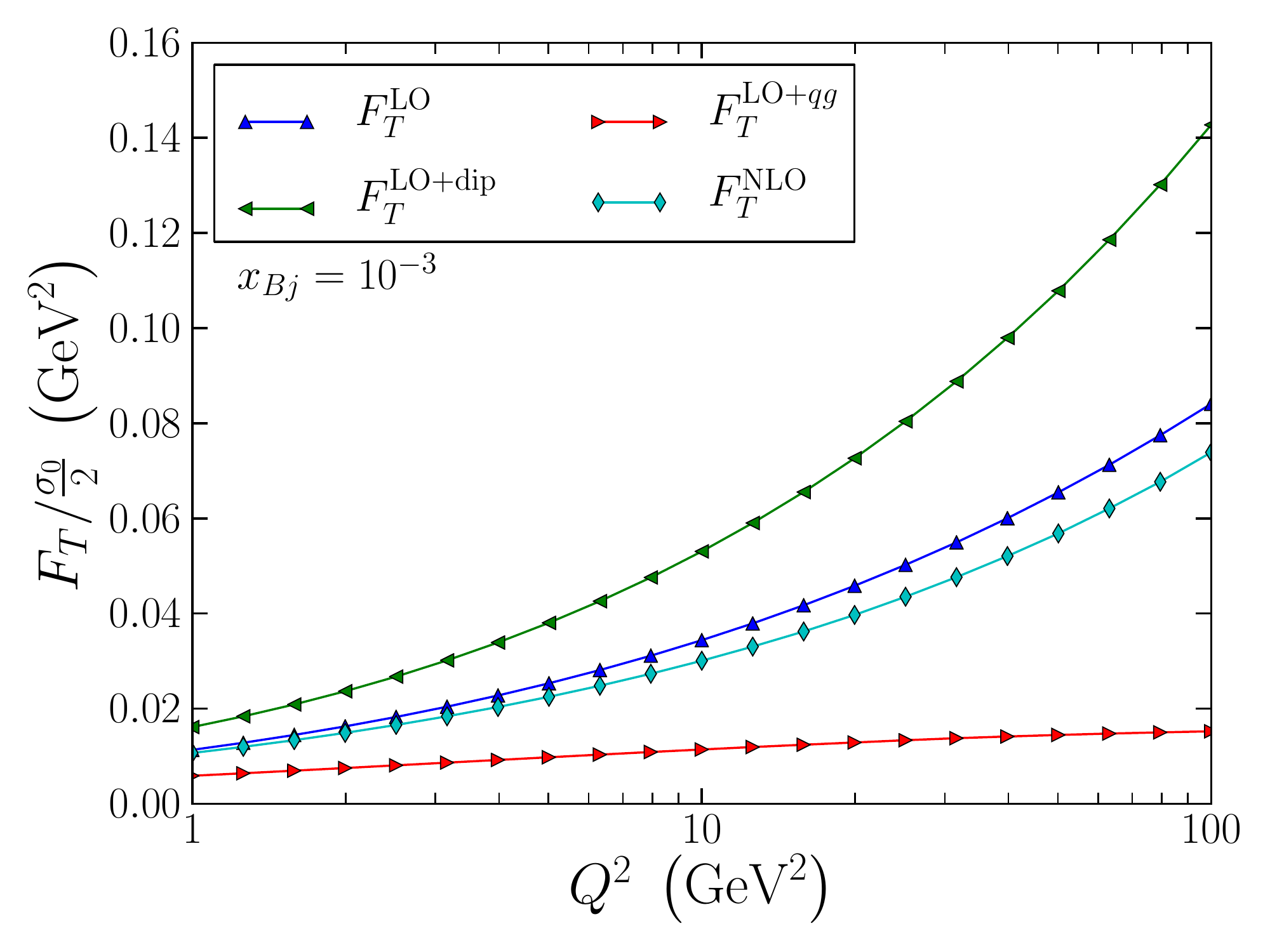}
	\caption{\footnotesize LO and NLO contributions to $F_L$ (left) and $F_T$ (right) as a function of $Q^2$ at $\xbj=10^{-3}$ with $\as=0.2$.}
	\label{fig:fc_Q}
\end{figure*}

In Fig. \ref{fig:fc_Q} we show the effect of the $\sigma^{qg}$ and $\sigma^{\text{dip}}$ NLO corrections to the structure functions as functions of $Q^2$ in the unsubtracted scheme \eqref{eq:NLO_qg_sub}. First we note that the NLO corrections overall are moderate and yield reasonable NLO results, similarly to the analogous scheme with single inclusive particle production \cite{Ducloue:2017mpb}. 
Overall the NLO corrections decrease the structure functions.

\begin{figure*}[tbp]
	\centering
	\includegraphics[scale=\figscale]{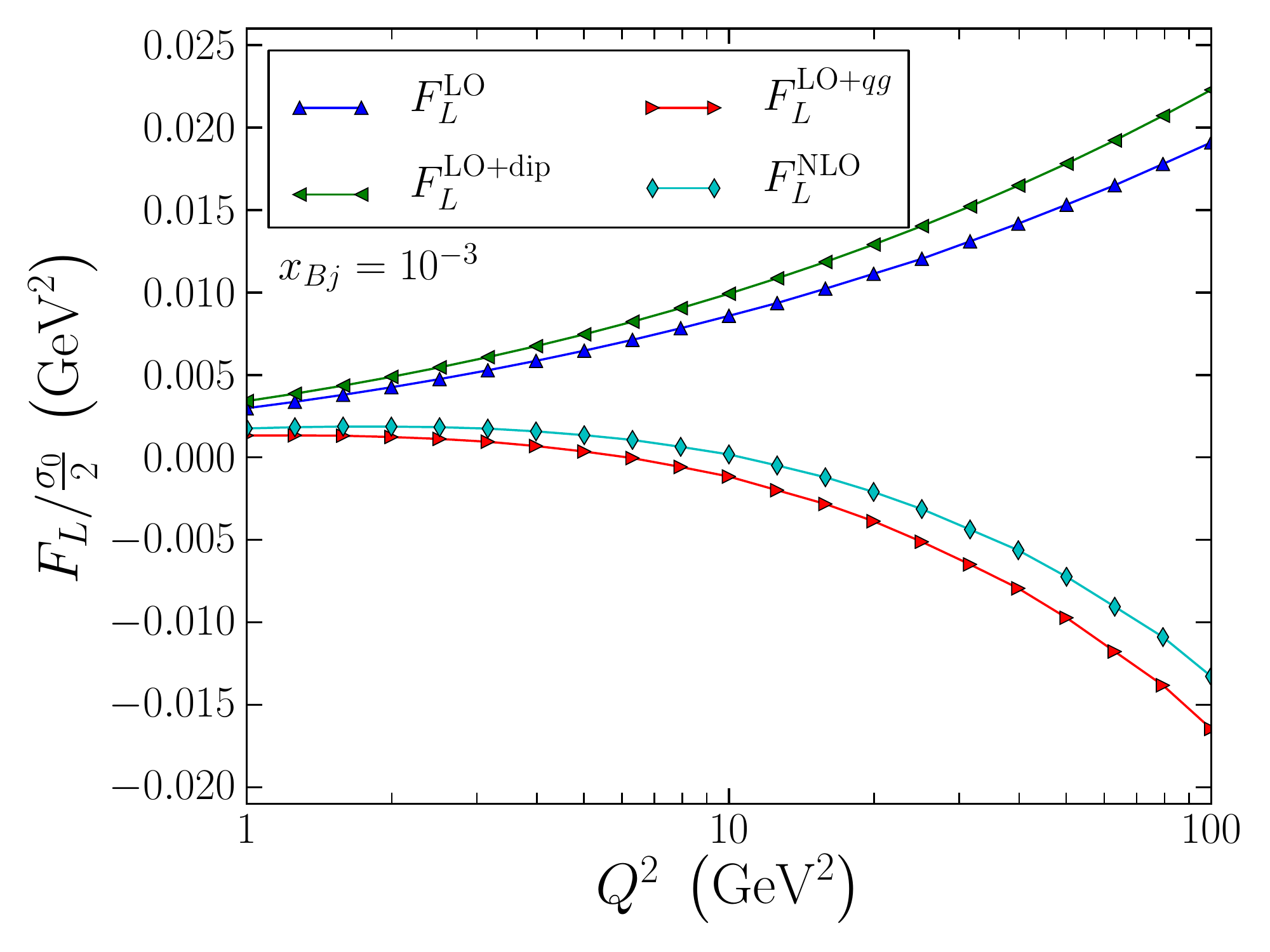}
	\hspace{\figspace}
	\includegraphics[scale=\figscale]{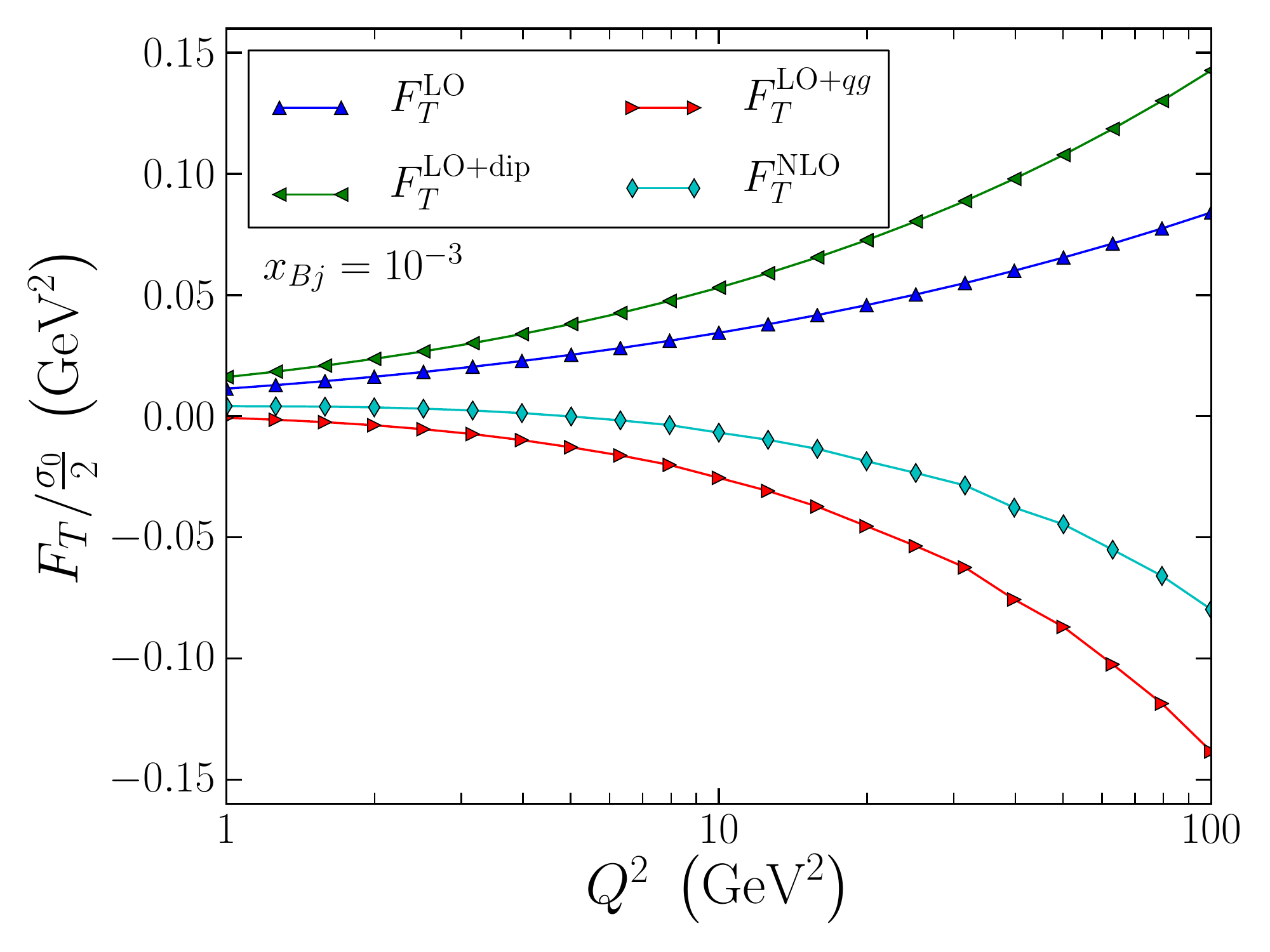}
	\caption{\footnotesize LO and NLO contributions to $F_L$ (left) and $F_T$ (right) as a function of $Q^2$ at $\xbj=10^{-3}$ with $\as=0.2$ and using the $\xbj$-subtraction procedure. }
	\label{fig:fc_Q_sub}
\end{figure*}

In Fig. \ref{fig:fc_Q_sub} we show the same quantities but with the $\xbj$-subtraction scheme \eqref{eq:NLO_xbjsub}. The quark-gluon contribution is again negative but even larger in magnitude, increasingly so at large $Q^2$, making the full NLO structure functions negative at $Q^2 \gtrsim 10 \, \text{GeV}^2$. So while the approximations made for the $\xbj$-subtraction scheme are in principle valid in a weak coupling sense, they have a large effect in practice in this region and can lead to unphysical results. This is similar to a negativity issue seen with single inclusive particle production at high transverse momenta \cite{Ducloue:2017mpb}.

\begin{figure*}[t]
	\centering
	\includegraphics[scale=\figscale]{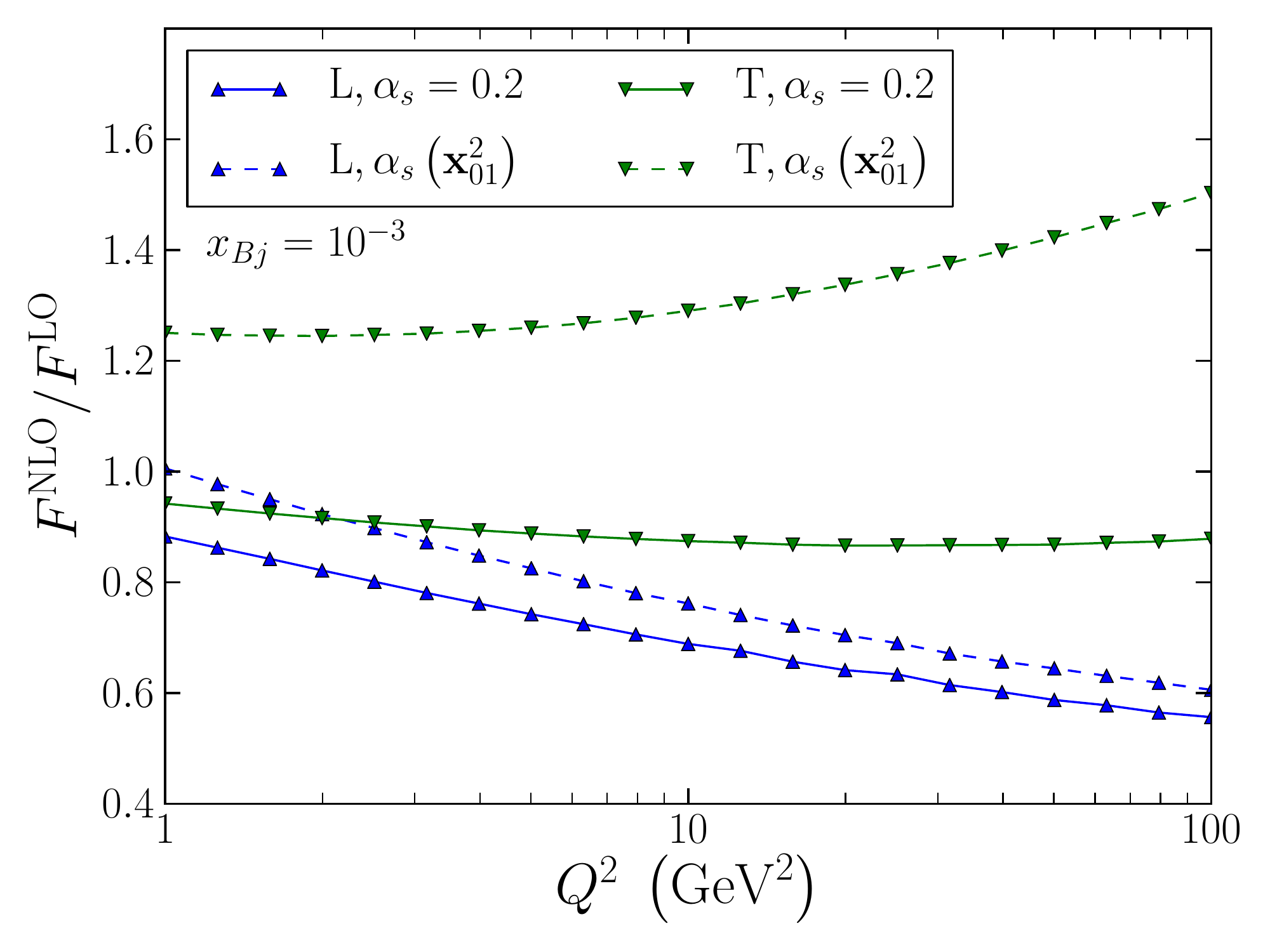}
	\hspace{\figspace}
	\includegraphics[scale=\figscale]{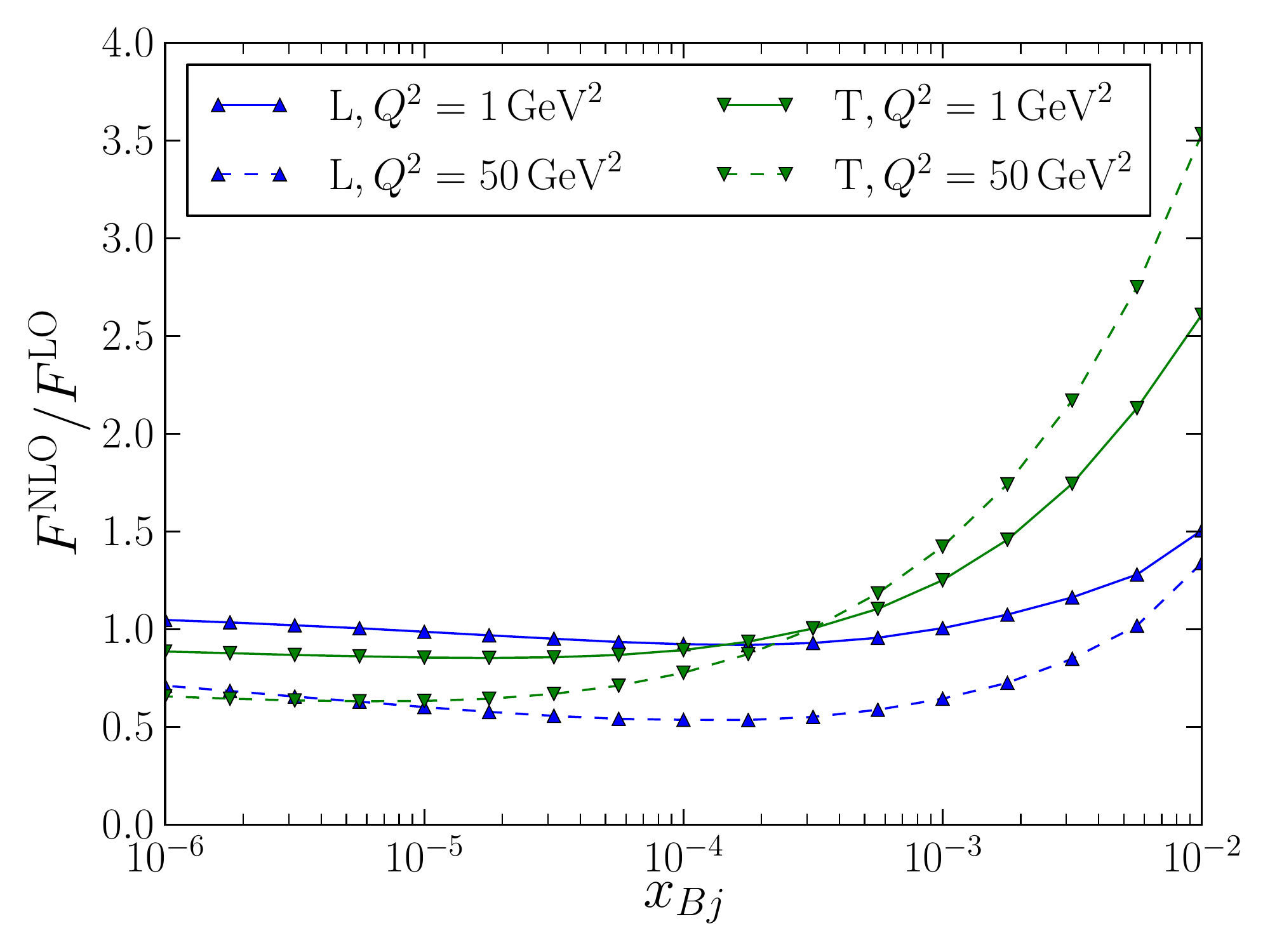}
	\caption{\footnotesize Left: NLO/LO ratio for $F_L$ and $F_T$ as a function of $Q^2$ at $\xbj=10^{-3}$ with fixed (solid) and running (dashed) coupling. Right: NLO/LO ratio for $F_L$ and $F_T$ as a function of $\xbj$ at $Q^2=1$ GeV$^2$ (solid) and $Q^2=50$ GeV$^2$ (dashed) with running coupling.}
	\label{fig:rc}
\end{figure*}

In Fig. \ref{fig:rc} we return to the working unsubtracted scheme to demonstrate the effects of the running coupling and the magnitude of the NLO corrections in a pair of ratio plots. In the left plot we see that the NLO corrections are of the order of a few tens of percent in the $Q^2$ range of interest. Secondly one should note that in this subtraction scheme the choice of running coupling can have large effects: the sign of the NLO correction to $F_T$ changes when one uses the parent dipole running coupling. This is due to the large cancellations between the different kinds of NLO contributions in this scheme. 
In the right plot the ratios with running coupling are shown as a function of $\xbj$ to demonstrate that in this scheme the NLO corrections increase in magnitude near the initial condition. While this transient effect, an artefact of the used subtraction scheme, does not interfere with the asymptotic high energy behavior, a careful treatment of this effect will be required for fits to data.

\section{Conclusions}
The recently calculated NLO corrections to DIS structure functions were made finite by a resummation of the large logarithms of energy and evaluated numerically. The NLO corrections were found to be of a reasonable magnitude in a working subtraction scheme, but sensitive to the details of the resummation and subtraction scheme. 
It was verified that the approximative subtraction scheme attempted in the past with single inclusive particle production fails to yield physical results in DIS at high $Q^2$ as well. A choice of the running coupling and a careful treatment of the discovered transient effect will be necessary in serious fits to HERA data.

\paragraph{Acknowledgments}
This work has been supported by the Academy of Finland, projects 273464 and 303756 and by the European Research Council, grant ERC-2015-CoG-681707.

\bibliographystyle{JHEP-2modM}
\bibliography{bib}

\ifx\mcitethebibliography\mciteundefinedmacro
\PackageError{JHEP-2modM.bst}{mciteplus.sty has not been loaded}
{This bibstyle requires the use of the mciteplus package.}\fi
\providecommand{\href}[2]{#2}
\begingroup\raggedright\begin{mcitethebibliography}{10}

\bibitem{Balitsky:1995ub}
I.~Balitsky,  \mbox{}
  \href{http://dx.doi.org/10.1016/0550-3213(95)00638-9}{{\em Nucl. Phys.} {\bf
  B463} (1996) 99} [\href{http://arXiv.org/abs/hep-ph/9509348}{{\tt
  arXiv:hep-ph/9509348 [hep-ph]}}]\relax
\mciteBstWouldAddEndPuncttrue
\mciteSetBstMidEndSepPunct{\mcitedefaultmidpunct}
{\mcitedefaultendpunct}{\mcitedefaultseppunct}\relax
\EndOfBibitem
\bibitem{Kovchegov:1999yj}
Y.~V. Kovchegov,  \mbox{}
  \href{http://dx.doi.org/10.1103/PhysRevD.60.034008}{{\em Phys. Rev.} {\bf
  D60} (1999) 034008} [\href{http://arXiv.org/abs/hep-ph/9901281}{{\tt
  arXiv:hep-ph/9901281 [hep-ph]}}]\relax
\mciteBstWouldAddEndPuncttrue
\mciteSetBstMidEndSepPunct{\mcitedefaultmidpunct}
{\mcitedefaultendpunct}{\mcitedefaultseppunct}\relax
\EndOfBibitem
\bibitem{Albacete:2010sy}
J.~L. Albacete {\em et.~al.},  \mbox{}
  \href{http://dx.doi.org/10.1140/epjc/s10052-011-1705-3}{{\em Eur. Phys. J.}
  {\bf C71} (2011) 1705} [\href{http://arXiv.org/abs/1012.4408}{{\tt
  arXiv:1012.4408 [hep-ph]}}]\relax
\mciteBstWouldAddEndPuncttrue
\mciteSetBstMidEndSepPunct{\mcitedefaultmidpunct}
{\mcitedefaultendpunct}{\mcitedefaultseppunct}\relax
\EndOfBibitem
\bibitem{Lappi:2013zma}
T.~Lappi and H.~Mäntysaari,  \mbox{}
  \href{http://dx.doi.org/10.1103/PhysRevD.88.114020}{{\em Phys. Rev.} {\bf
  D88} (2013) 114020} [\href{http://arXiv.org/abs/1309.6963}{{\tt
  arXiv:1309.6963 [hep-ph]}}]\relax
\mciteBstWouldAddEndPuncttrue
\mciteSetBstMidEndSepPunct{\mcitedefaultmidpunct}
{\mcitedefaultendpunct}{\mcitedefaultseppunct}\relax
\EndOfBibitem
\bibitem{Balitsky:2008zza}
I.~Balitsky and G.~A. Chirilli,  \mbox{}
  \href{http://dx.doi.org/10.1103/PhysRevD.77.014019}{{\em Phys. Rev.} {\bf
  D77} (2008) 014019} [\href{http://arXiv.org/abs/0710.4330}{{\tt
  arXiv:0710.4330 [hep-ph]}}]\relax
\mciteBstWouldAddEndPuncttrue
\mciteSetBstMidEndSepPunct{\mcitedefaultmidpunct}
{\mcitedefaultendpunct}{\mcitedefaultseppunct}\relax
\EndOfBibitem
\bibitem{Beuf:2014uia}
G.~Beuf,  \mbox{}  \href{http://dx.doi.org/10.1103/PhysRevD.89.074039}{{\em
  Phys. Rev.} {\bf D89} (2014) 074039}
  [\href{http://arXiv.org/abs/1401.0313}{{\tt arXiv:1401.0313 [hep-ph]}}]\relax
\mciteBstWouldAddEndPuncttrue
\mciteSetBstMidEndSepPunct{\mcitedefaultmidpunct}
{\mcitedefaultendpunct}{\mcitedefaultseppunct}\relax
\EndOfBibitem
\bibitem{Iancu:2015vea}
E.~Iancu {\em et.~al.},  \mbox{}
  \href{http://dx.doi.org/10.1016/j.physletb.2015.03.068}{{\em Phys. Lett.}
  {\bf B744} (2015) 293} [\href{http://arXiv.org/abs/1502.05642}{{\tt
  arXiv:1502.05642 [hep-ph]}}]\relax
\mciteBstWouldAddEndPuncttrue
\mciteSetBstMidEndSepPunct{\mcitedefaultmidpunct}
{\mcitedefaultendpunct}{\mcitedefaultseppunct}\relax
\EndOfBibitem
\bibitem{Iancu:2015joa}
E.~Iancu {\em et.~al.},  \mbox{}
  \href{http://dx.doi.org/10.1016/j.physletb.2015.09.071}{{\em Phys. Lett.}
  {\bf B750} (2015) 643} [\href{http://arXiv.org/abs/1507.03651}{{\tt
  arXiv:1507.03651 [hep-ph]}}]\relax
\mciteBstWouldAddEndPuncttrue
\mciteSetBstMidEndSepPunct{\mcitedefaultmidpunct}
{\mcitedefaultendpunct}{\mcitedefaultseppunct}\relax
\EndOfBibitem
\bibitem{Lappi:2016fmu}
T.~Lappi and H.~Mäntysaari,  \mbox{}
  \href{http://dx.doi.org/10.1103/PhysRevD.93.094004}{{\em Phys. Rev.} {\bf
  D93} (2016) 094004} [\href{http://arXiv.org/abs/1601.06598}{{\tt
  arXiv:1601.06598 [hep-ph]}}]\relax
\mciteBstWouldAddEndPuncttrue
\mciteSetBstMidEndSepPunct{\mcitedefaultmidpunct}
{\mcitedefaultendpunct}{\mcitedefaultseppunct}\relax
\EndOfBibitem
\bibitem{Beuf:2016wdz}
G.~Beuf,  \mbox{}  \href{http://dx.doi.org/10.1103/PhysRevD.94.054016}{{\em
  Phys. Rev.} {\bf D94} (2016) 054016}
  [\href{http://arXiv.org/abs/1606.00777}{{\tt arXiv:1606.00777
  [hep-ph]}}]\relax
\mciteBstWouldAddEndPuncttrue
\mciteSetBstMidEndSepPunct{\mcitedefaultmidpunct}
{\mcitedefaultendpunct}{\mcitedefaultseppunct}\relax
\EndOfBibitem
\bibitem{Beuf:2017bpd}
G.~Beuf,  \mbox{}  \href{http://dx.doi.org/10.1103/PhysRevD.96.074033}{{\em
  Phys. Rev.} {\bf D96} (2017) 074033}
  [\href{http://arXiv.org/abs/1708.06557}{{\tt arXiv:1708.06557
  [hep-ph]}}]\relax
\mciteBstWouldAddEndPuncttrue
\mciteSetBstMidEndSepPunct{\mcitedefaultmidpunct}
{\mcitedefaultendpunct}{\mcitedefaultseppunct}\relax
\EndOfBibitem
\bibitem{Hanninen:2017ddy}
H.~Hänninen, T.~Lappi and R.~Paatelainen,  \mbox{}
  \href{http://dx.doi.org/10.1016/j.aop.2018.04.015}{{\em Annals Phys.} {\bf
  393} (2018) 358} [\href{http://arXiv.org/abs/1711.08207}{{\tt
  arXiv:1711.08207 [hep-ph]}}]\relax
\mciteBstWouldAddEndPuncttrue
\mciteSetBstMidEndSepPunct{\mcitedefaultmidpunct}
{\mcitedefaultendpunct}{\mcitedefaultseppunct}\relax
\EndOfBibitem
\bibitem{Iancu:2016vyg}
E.~Iancu, A.~H. Mueller {\em et.~al.},  \mbox{}
  \href{http://dx.doi.org/10.1007/JHEP12(2016)041}{{\em JHEP} {\bf 12} (2016)
  041} [\href{http://arXiv.org/abs/1608.05293}{{\tt arXiv:1608.05293
  [hep-ph]}}]\relax
\mciteBstWouldAddEndPuncttrue
\mciteSetBstMidEndSepPunct{\mcitedefaultmidpunct}
{\mcitedefaultendpunct}{\mcitedefaultseppunct}\relax
\EndOfBibitem
\bibitem{Ducloue:2017mpb}
B.~Ducloué, T.~Lappi {\em et.~al.},  \mbox{}
  \href{http://dx.doi.org/10.1103/PhysRevD.95.114007}{{\em Phys. Rev.} {\bf
  D95} (2017) 114007} [\href{http://arXiv.org/abs/1703.04962}{{\tt
  arXiv:1703.04962 [hep-ph]}}]\relax
\mciteBstWouldAddEndPuncttrue
\mciteSetBstMidEndSepPunct{\mcitedefaultmidpunct}
{\mcitedefaultendpunct}{\mcitedefaultseppunct}\relax
\EndOfBibitem
\bibitem{Ducloue:2017ftk}
B.~Ducloué {\em et.~al.},  \mbox{}
  \href{http://dx.doi.org/10.1103/PhysRevD.96.094017}{{\em Phys. Rev.} {\bf
  D96} (2017) 094017} [\href{http://arXiv.org/abs/1708.07328}{{\tt
  arXiv:1708.07328 [hep-ph]}}]\relax
\mciteBstWouldAddEndPuncttrue
\mciteSetBstMidEndSepPunct{\mcitedefaultmidpunct}
{\mcitedefaultendpunct}{\mcitedefaultseppunct}\relax
\EndOfBibitem
\bibitem{McLerran:1993ni}
L.~D. McLerran {\em et.~al.},  \mbox{}
  \href{http://dx.doi.org/10.1103/PhysRevD.49.2233}{{\em Phys. Rev.} {\bf D49}
  (1994) 2233} [\href{http://arXiv.org/abs/hep-ph/9309289}{{\tt
  arXiv:hep-ph/9309289 [hep-ph]}}]\relax
\mciteBstWouldAddEndPuncttrue
\mciteSetBstMidEndSepPunct{\mcitedefaultmidpunct}
{\mcitedefaultendpunct}{\mcitedefaultseppunct}\relax
\EndOfBibitem
\end{mcitethebibliography}\endgroup

\end{document}